\documentclass[conference]{IEEEtran}

\IEEEoverridecommandlockouts

\usepackage{mathtools}
\usepackage{comment}
\usepackage{cite}
\usepackage{amssymb}
\usepackage{epsfig}
\usepackage{epsf}
\usepackage{amsmath,amsfonts,amsthm,bm} 
\usepackage{subfigure}
\usepackage{graphicx}
\usepackage{color}
\usepackage[noend]{algorithmic}
\usepackage{algorithm}
\usepackage{multicol}
\usepackage{tikz}
\usepackage{pgfplots}
\pgfplotsset{compat=1.7}
\usepgfplotslibrary{groupplots} 
\usetikzlibrary{pgfplots.groupplots}

\usepgfplotslibrary{fillbetween}

\usepackage{blindtext}

\usepackage{subfiles} 

\def\BibTeX{{\rm B\kern-.05em{\sc i\kern-.025em b}\kern-.08em
    T\kern-.1667em\lower.7ex\hbox{E}\kern-.125emX}}

\usepackage{url}

\usepackage[normalem]{ulem}

\newcommand{\newRIS}{I-RIS }
\newcommand{\newwRIS}{I-RIS}
\newcommand{\gidx}{k}
\begin{document}


\title{Mix-and-Conquer: Beamforming Design with
Interconnected RIS for Multi-User
Networks
\vspace{-4mm}











\thanks{The work of S. Nassirpour and A. Vahid was in part supported by the 2021 SONY Faculty Innovation Award and NSF grants: ECCS-2030285, CNS-2343959, CNS-2343964, and AST-2348589.}
}

\author{\IEEEauthorblockN{
Sajjad Nassirpour$^\ast$, Naoki Kusashima$^\dagger$, Jose Flordelis$^\dagger$, 
and Alireza Vahid$^\ddagger$}
\IEEEauthorblockA{\textit{$^\ast$Department of Electrical and Computer
Engineering, San Diego State University, San Diego, USA}\\
\textit{$^\dagger$Lund laboratory, Sony Europe B.V., Lund, Sweden}\\
\textit{$^\ddagger$Department of Electrical and Microelectronic Engineering, Rochester Institute of Technology, Rochester, USA}\\
snassirpour@sdsu.edu, naoki.kusashima@sony.com, jose.flordelis@sony.com, 
and alireza.vahid@rit.edu}
}
\vspace{-7mm}
\maketitle

\begin{abstract}






We propose a new reconfigurable intelligent surface (RIS) structure, referred to as interconnected RIS (I-RIS), which allows the RIS elements to be interconnected and share the incident signals using simple binary radio frequency (RF) switches and mix them into the reflecting signals. This structure enables multi-user scaling and requires fewer elements (i.e., a compact structure) compared to standard RIS (S-RIS), which assumes no interconnection between the elements. The I-RIS compact design makes it practical for deployment on space-limited nodes, e.g., unmanned aerial vehicles (UAVs). Hence, in this work, we propose a beamforming design based on I-RIS in a multi-user network, where we use binary RF switches as RIS elements. We show that our switch-based I-RIS offers a higher gain compared to an S-RIS using phase shifters. Finally, we introduce two optimization methods, sigmoid filled function (SFF) and semi-definite binary optimization (SBO), to optimize the RIS elements and evaluate their performance in terms of sum-rate and complexity.
\end{abstract}

\section{Introduction}
\label{Section:intro}
With the ever-increasing demand for wireless networks, the next generation of wireless networks must improve key performance metrics such as coverage, data rate, and spectrum/energy efficiency compared to the existing networks. Reconfigurable intelligent surface (RIS) has recently emerged as a promising technique, offering a range of advantages, including the capability to bypass obstacles to improve coverage and enhance data rate and spectral efficiency, all achieved in a cost-effective and power-efficient manner~\cite{zhang2020capacity,yu2019miso,bjornson2020reconfigurable}. RIS, an artificial surface comprising electromagnetic material, contains a substantial number of reflecting elements that are low-cost and (nearly-)passive, capable of independently adjusting the phase and/or amplitude of incident signals. By intelligently configuring the RIS elements, it becomes possible to combine reflected signals constructively at the intended receiver and destructively at undesired receivers.


Due to high path loss in the transmitter-RIS-receiver link and the passive nature of RIS, existing RISs, referred to as standard RIS (S-RIS), rely on many elements for network assistance; each S-RIS element acts individually, with no interconnection between elements, leading to a diagonal RIS matrix. However, achieving a large-size RIS is often impractical, especially in space-limited nodes, e.g., unmanned aerial vehicles (UAVs), where space significantly influences weight, power consumption, and vehicle cost. Minimizing component size and using space efficiently allows UAVs to enhance payload/fuel capacity, agility, and extend flight times~\cite{ramasamy2014avionics,azarov2019composite}.

As discussed above, small-size (i.e., compact) RIS is more desired in space-limited nodes. Thus, in this paper, we propose a new RIS structure, referred to as interconnected RIS (I-RIS), where RIS elements have interconnection with each other, leading to a non-diagonal I-RIS matrix. A study in~\cite{an2023stacked} proposed a single-user stacked intelligent metasurface (SIM)-assisted network, which also generates a non-diagonal matrix. 
However, our focus in this paper is on a multi-user scenario. 
In I-RIS, we divide the elements into different cells and define the elements in each cell as {\it neighbors}. Recently,~\cite{li2022reconfigurable} studied an I-RIS, where each element receives the incident signal and reflects it either via itself or via only one of the other elements. However, in our work, each I-RIS element involves: (i) sharing its incident signal with multiple of its neighbors and (ii) reflecting a combination of the incident signals from its neighbors toward the receivers. We show that I-RIS improves user-scaling and requires fewer elements compared to S-RIS. 
For example, our findings demonstrate that in a single-input single-output (SISO) network serving four users, an I-RIS with $64$ elements and $16$ cells provides an $81 \%$ higher sum-rate than an S-RIS with the same number of elements. Additionally, I-RIS requires four times fewer elements to match the same sum-rate achieved by S-RIS. The smaller size of I-RIS is feasible due to what we call the degrees-of-control (DoC), which captures the number of tunable parameters in RIS and intuitively determines the gain from incorporating an RIS into the system. Thanks to the interconnected design, I-RIS needs fewer elements to get the same DoC than prior designs.



Beyond the issue of interconnected elements, previous works usually used phase shifters (PSs) as RIS elements. However, simultaneously controlling the amplitude and phase of PSs is challenging. 
Motivated by this,~\cite{mendez2016hybrid,nassirpour2022power} proposed beamforming techniques using radio frequency (RF) switches to enhance performance. Consequently, we consider RF switches as RIS elements in this work and show that our switch-based I-RIS outperforms an S-RIS using low-bit PSs.
On the other hand, the use of RF switches introduces a non-convex integer programming problem. To address this problem, we propose two optimization methods: the sigmoid filled function (SFF) method and the semi-definite binary optimization (SBO) method, where the former is a heuristic approach, and the latter utilizes the successive convex approximation (SCA) method. We compare the performance of these two optimization methods with the successive-refinement (SR) method~\cite{wu2019beamforming} in terms of sum-rate and complexity. Our findings reveal: (i) the SFF method provides a higher sum-rate than the others; (ii) the complexity of the SFF method is lower than the SBO method and higher than the SR method.

The rest of this paper is organized as follows:
Section~\ref{Section:Problem} demonstrates problem setting and RIS structure. Section~\ref{Section:Beam_design} presents our beamforming
design, Section~\ref{Section:Num_results} includes numerical analysis, and Section~\ref{Section:conclusion} concludes the paper.

\textbf{Notation:} We utilize italic, bold-face lowercase, and bold-face uppercase letters to represent scalars, vectors, and matrices, respectively. We use $\mathbb{C}^{a\times b}$ to denote the space of an $a\times b$ dimensional complex-valued matrix. $\mathrm{diag}\{\textbf{a}\}$ shows a diagonal matrix using $\textbf{a}$, $\textbf{Y} \succeq \mathbf{0}$ denotes $\textbf{Y}$ is a positive semi-definite matrix, $\mathrm{blkdiag}\{\mathbf{A}, \mathbf{B}\}$ is a block diagonal matrix with $\mathbf{A}$ and $\mathbf{B}$ on its main diagonal, and $\mathrm{Diag}(\textbf{A})$ is a vector of diagonal elements in $\textbf{A}$. Here, $||\textbf{A}||_{\mathrm{F}}$ denotes the Frobenius norm of $\textbf{A}$, $A_{i,j}$ is the element in the $i^{\mathrm{th}}$ row and $j^{\mathrm{th}}$ column of $\textbf{A}$, $|a|$ is the modulus of $a$, and $||\textbf{a}||$ denotes Euclidean norm of $\textbf{a}$. Further, we use $\log(.)$, $\lfloor . \rfloor$,  $\mathrm{tr}\left(.\right)$, $\textbf{B}^{\top}$, and $\textbf{B}^H$ to indicate the logarithmic function in base 2, the floor function, the trace function, the transpose of $\textbf{B}$, and the conjugate transpose of $\textbf{B}$, respectively. We also denote sets using calligraphic font style, and for a set $\mathcal{M}$, $|\mathcal{M}|$ is its cardinality.

\section{Problem Setting and RIS Structure}
\label{Section:Problem}
\begin{figure}[ht]
\vspace{-3mm}
  \centering
  \includegraphics[trim = 0mm 0mm 0mm 0mm, clip, scale=7, width=0.75\linewidth, draft=false]{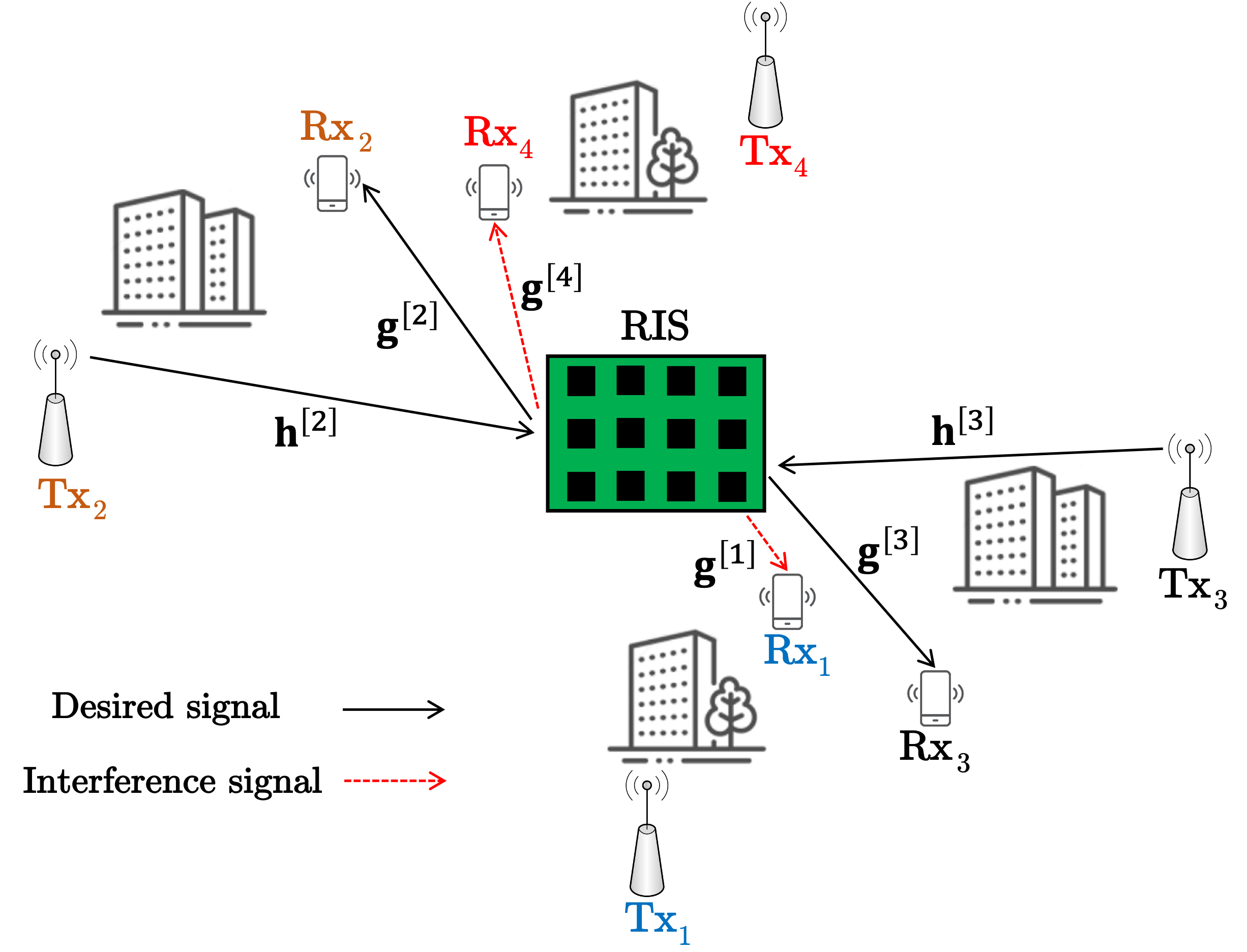}
  \vspace{-4mm}
  \caption{A SISO network with an RIS assisting multi-user communications.
}\label{Fig:channel_model_standard_RIS}
  \vspace{-2mm}
\end{figure}

In this paper, we examine an RIS-assisted network with distributed transmitter-receiver pairs where each transmitter aims to communicate with only one desired receiver, as shown in Fig.~\ref{Fig:channel_model_standard_RIS}. Further, we focus on a SISO scenario to highlight the pure RIS gain in the network.

\subsection{Channel Model} We denote the channels between ${\sf Tx}_j$ and RIS, and between RIS and ${\sf Rx}_{k}$ at time $t$ as $\textbf{h}^{j}(t) \in \mathbb{C}^{M\times 1}$ and $\textbf{g}^{k}(t) \in \mathbb{C}^{1\times M}$, respectively, where $j,k \in \{1,2,\ldots,K\}$, $K$ is the number of transmitter-receiver pairs, and $M$ is the number of RIS elements. We use Rician fading to model $\textbf{h}^{j}(t)$ and express $h_m^{j}(t)$, the $m^{\mathrm{th}}$ element of $\textbf{h}^{j}(t), 1 \leq m \leq M$, as follows:
\begin{align}
    \label{eq:Rician_fading}
    h_m^{j}(t)=\sqrt{\frac{1}{1+\kappa}}h_m^{j,NLoS}(t)+\sqrt{\frac{\kappa}{1+\kappa}}h_m^{j,LoS},
\end{align}
where $\kappa$ is the Rician factor, $h_m^{j,NLoS}(t) \thicksim \mathcal{CN}(0,C_0d_{1,j}^{-\alpha_1})$ represents the non-line-of-sight (NLoS) component of $h_m^{j}(t)$, and $h_m^{j,LoS}=\sqrt{C_0d_{1,j}^{-\alpha_1}}e^{-j\frac{2 \pi }{\lambda}d_{1,j}}$ shows the line-of-sight (LoS) part of $h_m^{j}(t)$, where $d_{1,j}$ denotes the distance between ${\sf Tx}_j$ and RIS, $\alpha_1$ is the path loss exponent between them, and $\lambda$ is the wavelength of the transmitted signal. We use $C_0$ to capture the signal loss at a reference distance, e.g., $1\mathrm{m}$. Here, $\textbf{g}^{k}(t)$ follows Rayleigh fading whose elements are modeled as a complex Gaussian distribution with zero mean and variance $C_0d_{2,k}^{-\alpha_2}$, where $d_{2,k}$ and $\alpha_2$ denote the distance and path loss exponent between RIS and ${\sf Rx}_k$, respectively. 

We assume that the direct path between transmitters and receivers is blocked. We also suppose that the channels are independent due to the spacing of the RIS elements being at least $\lambda/2$ apart. Similar to~\cite{fu2021reconfigurable,wu2019intelligent}, we assume that channel state information (CSI) is available globally, and beyond that, there is no data exchange between transmitters or receivers. We note that all operations occur within a single coherence time; hence, we omit the time notation in the rest of this paper.

\begin{figure*}[ht]
  \centering
  \includegraphics[trim = 0mm 0mm 0mm 0mm, clip, scale=4, width=0.85\linewidth, draft=false]{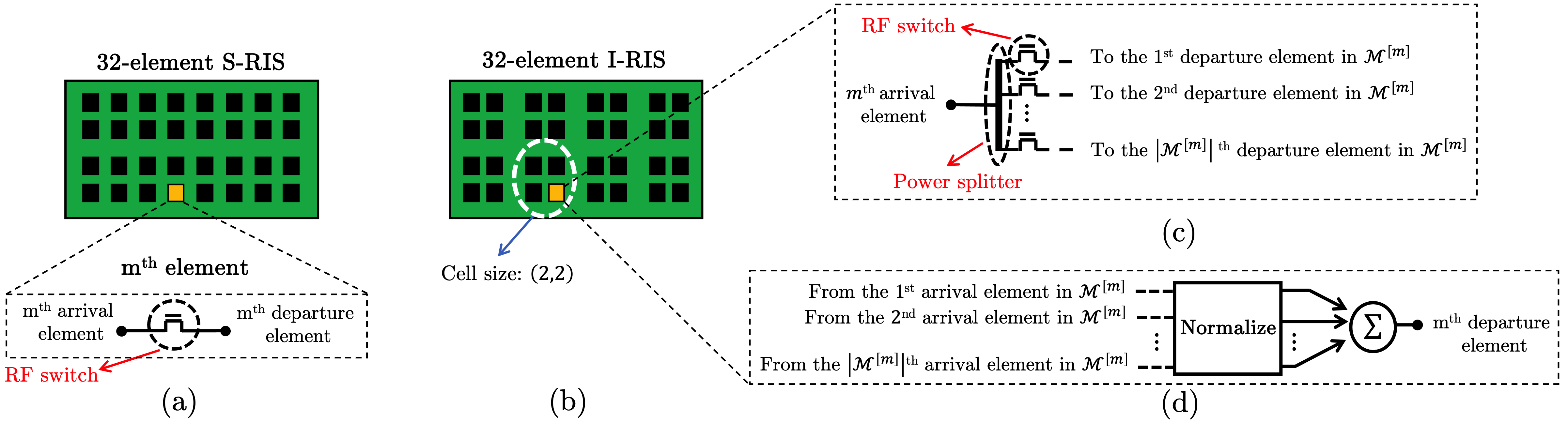}
  \vspace{-4mm}
  \caption{(a) A $32$-element S-RIS, where the $m^{\mathrm{th}}$ RIS element uses an RF switch to either reflect or block its incident signal; (b) An I-RIS with $32$ elements and eight cells of size $(2,2)$; (c) The $m^{\mathrm{th}}$ arrival element uses $cd$ RF switches and a power splitter to share its incident signal; (d) The reflected signal from the $m^{\mathrm{th}}$ departure element includes a combination of the incident signals at all arrival elements.}\label{Fig:channel_model_new_RIS}
  \vspace{-2mm}
\end{figure*}
\subsection{Existing vs. The Proposed RIS Design}
In this subsection, we delve into the details of two RIS structures using RF switches. Here, each RIS element has an arrival part that receives the incident signals from transmitters and a departure part that reflects the signals toward receivers.

\textbf{S-RIS:} In this RIS, there is no interconnection between the elements. The $m^{\mathrm{th}}$ RIS element, $1 \leq m \leq M$, receives the incident signal and then uses its binary RF switch, $s_{m}\in \{0,1\}$ between its arrival and departure parts, to either reflect or block its incident signal as shown in Fig.~\ref{Fig:channel_model_new_RIS}(a). We define $\mathbf{T}^{\mathrm{s}}$ as the S-RIS configuration, which is given by:
\begin{align}
\mathbf{T}^{\mathrm{s}}=\mathrm{diag}\left\{\boldsymbol{\tau}^{\mathrm{s}}\right\},~~~\boldsymbol{\tau}^{\mathrm{s}}=\left[s_{1},s_{2},\ldots,s_{M}\right]^{\top}.
\end{align}

\textbf{I-RIS:} Here, we first divide RIS into $M/(c \times d)$ different cells, where $c$ and $d$ represent the cell size. (e.g., Fig.~\ref{Fig:channel_model_new_RIS}(b) shows a $32$-element I-RIS with eight cells of size $(2, 2)$.) Without loss of generality, we suppose $M$ is divisible by $c \times d$. In the rest of this work, we use $cd$ to denote $c \times d$. Next, each element within a given cell shares its incident signal with its neighbors in that cell, as depicted in Fig.~\ref{Fig:channel_model_new_RIS}(c). We define $\mathcal{M}^{[m]}$ as a cell that includes the $m^{\mathrm{th}}$ element. Then, the $m^{\mathrm{th}}$ RIS element uses $cd$ binary RF switches to determine  $\bar{\mathcal{M}}^{[m]}\subseteq \mathcal{M}^{[m]}$, which contains the active elements in the cell obtaining the incident signal received by the $m^{\mathrm{th}}$ arrival element. The $m^{\mathrm{th}}$ arrival element uses a power splitter to distribute the incident signal power uniformly across all members of $\bar{\mathcal{M}}^{[m]}$. Then, to indicate how the $m^{\mathrm{th}}$ arrival element connects to the $\ell^{\mathrm{th}}$ departure element, $1\leq \ell,m\leq M$, we define $\tau_{\ell, m}$ as:
$\tau_{\ell, m}=\left(1/\sqrt{\left|\bar{\mathcal{M}}^{[m]}\right|}\right)s_{\ell, m}$ when $\ell\in \mathcal{M}^{[m]}$; conversely, $\tau_{\ell, m}=0$ if $\ell\notin \mathcal{M}^{[m]}$.
Here, $s_{\ell, m}\in \{0,1\}$ is a binary RF switch between the $m^{\mathrm{th}}$ arrival element and the $\ell^{\mathrm{th}}$ departure element, and $1/\sqrt{\left|\bar{\mathcal{M}}^{[m]}\right|}$ guarantees $\sum_{\ell=1}^M |\tau_{\ell, m}|^2\leq1$, where the inequality holds true only if all switches from the $m^{\mathrm{th}}$ arrival element to its neighbors are zero. In this case $\sum_{\ell=1}^M |\tau_{\ell, m}|^2 =0 <1$. 
Then, we use $\mathbf{T}^{\mathrm{i}}$ to denote the I-RIS configuration as:
\begin{align}
    \mathbf{T}^{\mathrm{i}}=\mathrm{blkdiag}\left\{\boldsymbol{\tau}^{\mathrm{i}[1]},\boldsymbol{\tau}^{\mathrm{i}[2]}, \ldots, \boldsymbol{\tau}^{\mathrm{i}[\frac{M}{cd}]}\right\},
\end{align}
where $\boldsymbol{\tau}^{\mathrm{i}[u]}, 1\leq u \leq \frac{M}{cd}$ follows \eqref{eq:tau_general_vector}.
\begin{figure*}
\begin{align}
\label{eq:tau_general_vector}
    \boldsymbol{\tau}^{\mathrm{i}[u]} = \begin{bmatrix}
\tau^{\mathrm{i}[u]}_{1,1} & \tau^{\mathrm{i}[u]}_{1,2} & \ldots & \tau^{\mathrm{i}[u]}_{1,cd}\\
\vdots & \vdots & \ldots & \vdots
\\
\tau^{\mathrm{i}[u]}_{cd,1} & \tau^{\mathrm{i}[u]}_{cd,2} & \ldots & \tau^{\mathrm{i}[u]}_{cd, cd}
\end{bmatrix} = \begin{bmatrix}
\tau_{(u-1)cd+1,(u-1)cd+1} & \tau_{(u-1)cd+1,(u-1)cd+2} & \ldots & \tau_{(u-1)cd+1,ucd}\\
\vdots & \vdots & \ldots & \vdots
\\
\tau_{ucd,(u-1)cd+1} & \tau_{ucd,(u-1)cd+2} & \ldots & \tau_{ucd,ucd}
\end{bmatrix}.
\end{align}
\hrule
\end{figure*}
We consider $\boldsymbol{\tau}^{\mathrm{i}[u]}=\hat{\boldsymbol{\tau}}^{\mathrm{i}[u]}+\Tilde{\boldsymbol{\tau}}^{\mathrm{i}[u]}$. For each $\tau^{\mathrm{i}[u]}_{u_1,u_2}, 1 \leq u_1, u_2 \leq cd$, if it is the only non-zero element in the $u_1^{\mathrm{th}}$ row and $u_2^{\mathrm{th}}$ column of $\boldsymbol{\tau}^{\mathrm{i}[u]}$, we include $\tau^{\mathrm{i}[u]}_{u_1,u_2}$ in $\hat{\boldsymbol{\tau}}^{\mathrm{i}[u]}$ by following~\eqref{eq:hat_tau}; otherwise, we save $\tau^{\mathrm{i}[u]}_{u_1,u_2}$ in $\Tilde{\boldsymbol{\tau}}^{\mathrm{i}[u]}$ as~\eqref{eq:tilde_tau}. We show \eqref{eq:tau_general_vector},~\eqref{eq:hat_tau}, and~\eqref{eq:tilde_tau} at the top of the next page.  
\begin{figure*}
\begin{minipage}[t]{0.50\linewidth}
\begin{equation}
\label{eq:hat_tau}
\hat{\tau}^{\mathrm{i}[u]}_{u_1,u_2}=\left\{ \begin{array}{ll} \tau^{\mathrm{i}[u]}_{u_1,u_2}, & \sum_{q=1}^{cd}\left(\tau^{\mathrm{i}[u]}_{u_1,q} + \tau^{\mathrm{i}[u]}_{q,u_2} \right) = 2\tau^{\mathrm{i}[u]}_{u_1,u_2}, \\
0, & \text{otherwise},
\end{array} \right.
\end{equation}
\end{minipage}
\begin{minipage}[t]{0.50\linewidth}
\begin{equation}
\label{eq:tilde_tau}
\tilde{\tau}^{\mathrm{i}[u]}_{u_1,u_2}=\left\{ \begin{array}{ll} \tau^{\mathrm{i}[u]}_{u_1,u_2}, & \sum_{q=1}^{cd}\left(\tau^{\mathrm{i}[u]}_{u_1,q} + \tau^{\mathrm{i}[u]}_{q,u_2}\right) \neq 2\tau^{\mathrm{i}[u]}_{u_1,u_2}, \\
0, & \text{otherwise},
\end{array} \right.
\end{equation}
\end{minipage}
\hrule
\end{figure*}Then, based on Fig.~\ref{Fig:channel_model_new_RIS}(d), to ensure I-RIS is passive, we normalize $\Tilde{\boldsymbol{\tau}}^{\mathrm{i}[u]}$, i.e., replace $\Tilde{\boldsymbol{\tau}}^{\mathrm{i}[u]}$ with $\Tilde{\boldsymbol{\tau}}^{\mathrm{i}[u]}/||\Tilde{\boldsymbol{\tau}}^{\mathrm{i}[u]}||_{\mathrm{F}}$. Finally, the $m^{\mathrm{th}}$ departure element mixes the incident signals of all elements in $\mathcal{M}^{[m]}$ and reflects a combination of them toward the receivers.

\textbf{Received signal:} ${\sf Rx}_{k}, 1\leq k \leq K,$ receives the reflected signals from the RIS, which is given by:
\begin{align}
\label{eq_ch5:received_sig_a_RIS}
    y_{k}=\sum_{j=1}^K \Big[\textbf{g}^{k}\mathbf{T}\textbf{h}^{j}\Big]x_j+n_{k},
\end{align}
where $\mathbf{T} \in \left\{\mathbf{T}^{\mathrm{s}}, \mathbf{T}^{\mathrm{i}}\right\}$, $x_j$ is the transmitted signal from ${\sf Tx}_j$, and $n_{k} \thicksim \mathcal{CN}(0,\sigma^2)$ denotes noise at ${\sf Rx}_{k}$. Here, $\mathbb{E}\left\{\textbf{x}^{H}\textbf{x}\right\}=\mathrm{diag}([P_1, P_2,\ldots,P_K])$, where $\textbf{x}=[x_1,x_2,\ldots,x_K] \in \mathbb{C}^{1\times K}$, and $P_j$ is the transmit power from ${\sf Tx}_j$.  Then, we use \eqref{eq_ch5:received_sig_a_RIS} to express the signal-to-interference-plus-noise ratio (SINR) at ${\sf Rx}_{k}$ as follows:
\begin{align}
\label{eq_ch5:SINR_update_a_RIS}
    \text{SINR}_{k}=\frac{P_{k}\Big|\textbf{g}^{k}\mathbf{T}\textbf{h}^{k}\Big|^2}{\sigma^2+\sum_{j=1, j\neq k}^{K }P_j\Big|\textbf{g}^{k}\mathbf{T}\textbf{h}^{j}\Big|^2}.
\end{align} 

\section{Beamforming Design}
\label{Section:Beam_design}
We aim to optimize RIS elements to maximize the sum-rate; thus, we formulate our optimization problem as:
\begin{align}
\label{eq:opt_prob_std_RIS_feb_27_2023}
    &\underset{\mathbf{T}}{\max}\underbrace{\sum_{\gidx=1}^{K} \log(1+\text{SINR}_{\gidx})}_{\overset{\triangle}=~-w(\mathbf{T})}\equiv \underset{\mathbf{T}}{\min}~w(\mathbf{T})\\ \nonumber
    &\text{s.t.} ~~\mathbf{T} \in \{\mathbf{T}^{\mathrm{s}}, \mathbf{T}^{\mathrm{i}}\},~~  s_m,s_{\ell, m}\in\{0,1\},~~1\leq \ell,m\leq M,
\end{align}
which is non-convex due to the binary constraints and the non-convex objective function. To optimize~\eqref{eq:opt_prob_std_RIS_feb_27_2023}, we first focus on S-RIS where $\mathbf{T}=\mathbf{T}^{\mathrm{s}}$ and propose two methods to solve the problem: (1) we use the 
semi-definite binary optimization (SBO) method, which converts~\eqref{eq:opt_prob_std_RIS_feb_27_2023} to a semi-definite relaxation (SDR) problem and then solves the equivalent convex problem; (2) we utilize the sigmoid filled function (SFF), which is a heuristic method based on filled function that directly targets \eqref{eq:opt_prob_std_RIS_feb_27_2023} to obtain an approximation of the global optimum solution. We provide further details below. 

\textbf{SBO method:} In this case, we use slack variable $\textbf{S}$ and rewrite the binary constraint as follows:

\begin{minipage}[ht]{0.5\linewidth}
\begin{equation}
\label{eq:Diag_S_feb_27_2023}
    \mathrm{Diag}(\textbf{S})=\boldsymbol{\tau}^{\mathrm{s}}, \quad \quad \quad
\end{equation}
\end{minipage}
\begin{minipage}[ht]{0.44\linewidth}
\begin{equation}
\label{eq:S=ss_feb_27_2023}
 \textbf{S}=\boldsymbol{\tau}^{\mathrm{s}}{\boldsymbol{\tau}^{\mathrm{s}}}^{\top},\quad \quad
\end{equation}
\end{minipage}

\begin{align}
\label{eq:01_values_feb_27_2023}
    0 \leq s_m \leq 1,~\text{and}~0 \leq S_{\ell, m} \leq 1,~~ \text{for}~ 1\leq \ell,m \leq M,
\end{align}
where $S_{\ell, m}$ is the element in the $\ell^{\mathrm{th}}$ row and $m^{\mathrm{th}}$ column of $\textbf{S}$. Here, \eqref{eq:S=ss_feb_27_2023} is non-convex; therefore, we substitute it with the following conditions~\cite{guo2018new,hu2021robust}.
\begin{align}
\label{eq:Y_matrix_feb_27_2023}
    \textbf{Y}=\begin{bmatrix}
        \textbf{S}& \boldsymbol{\tau}^{\mathrm{s}}\\
        {\boldsymbol{\tau}^{\mathrm{s}}}^{\top}& 1\\
    \end{bmatrix}\succeq \mathbf{0},~~\text{and} ~~ \text{Rank}(\textbf{Y})=1.
\end{align}

Since $\textbf{Y}$ includes binary elements, the rank of $\textbf{Y}$ is equal to one for any binary feasible solution. Next, we define $\textbf{G}^{\gidx j}\overset{\triangle}=\textbf{g}^{\gidx} \mathrm{diag}(\textbf{h}^{j})$ and $\bar{\textbf{G}}^{\gidx j}\overset{\triangle}=\textbf{G}^{\gidx j}{\textbf{G}^{\gidx j}}^{\textrm{H}}$ and use \eqref{eq:Diag_S_feb_27_2023}, \eqref{eq:01_values_feb_27_2023}, and \eqref{eq:Y_matrix_feb_27_2023} to rewrite~\eqref{eq:opt_prob_std_RIS_feb_27_2023} as an SDR problem as follows:
\begin{align}
\label{eq:finalswitch_SNR_feb_27_2023}
    &\underset{\boldsymbol{\tau}^{\mathrm{s}},\textbf{S} }{\min}~-\sum_{\gidx=1}^K\log\left(1+\frac{P_{\gidx}~\mathrm{tr}\left(\textbf{S}\bar{\textbf{G}}^{\gidx \gidx}\right)}{\sigma^2+\sum_{j=1,j\neq \gidx}^K P_j~\mathrm{tr}\left(\textbf{S}\bar{\textbf{G}}^{\gidx j}\right) }\right)\\ \nonumber
    &\text{s.t.} ~~ \mathrm{Diag}(\textbf{S})=\boldsymbol{\tau}^{\mathrm{s}}, \quad \textbf{Y}\succeq \mathbf{0}, \\ \nonumber 
    & ~~~~~~0 \leq s_m \leq 1,~~0 \leq S_{\ell, m} \leq 1,~~ \text{for}~ 1\leq \ell,m \leq M.
\end{align}

Then, we use the feature of a logarithmic function and rewrite~\eqref{eq:finalswitch_SNR_feb_27_2023} as:
\begin{align}
\label{eq:finalswitch_SINR_Dbi_Nbi_feb_27_2023}
    &\underset{\boldsymbol{\tau}^{\mathrm{s}},\textbf{S} }{\min}~N-D\\ \nonumber
    &\text{s.t.} ~~ \mathrm{Diag}(\textbf{S})=\boldsymbol{\tau}^{\mathrm{s}}, \quad \textbf{Y}\succeq \mathbf{0}, \\ \nonumber 
    & ~~~~~~0 \leq s_m \leq 1,~~0 \leq S_{\ell, m} \leq 1,~~ \text{for}~ 1\leq \ell,m \leq M,
\end{align}
where
\begin{align}
    \label{eq:SINR_noise_Nbi_feb_27_2023}
    N=-\sum_{\gidx=1}^K\log\left(\sigma^2+\sum_{j=1}^K P_{j}~\mathrm{tr}\left(\textbf{S}\bar{\textbf{G}}^{\gidx j}\right)\right),
\end{align}
and
\begin{align}
    \label{eq:SINR_noise_Dbi_feb_27_2023}
    D=-\sum_{\gidx=1}^K\log\left(\sigma^2+\sum_{j=1, j\neq \gidx}^K P_j~\mathrm{tr}\left(\textbf{S}\bar{\textbf{G}}^{\gidx j}\right)\right).
\end{align}

Based on \eqref{eq:SINR_noise_Nbi_feb_27_2023} and \eqref{eq:SINR_noise_Dbi_feb_27_2023}, $N$ and $D$ are convex with respect to $\textbf{S}$; thus,~\eqref{eq:finalswitch_SINR_Dbi_Nbi_feb_27_2023} falls under the difference of convex (DC) optimization problem. To solve this problem, we use the SCA method, which is an iterative approach that runs for $I_{\mathrm{tr}}$ times. To do so, we apply the first-order Taylor approximation to $D$ and obtain a lower bound on that as follows:
\begin{align}
\label{eq:lower_bound_Dbi_formula_feb_27_2023}
    D(\textbf{S})\geq D\left(\textbf{S}^t\right)+\nabla^{\top}_{\textbf{S}}\left(D\left(\textbf{S}^t\right)\right)\left(\textbf{S}-\textbf{S}^t\right),
\end{align}
where $\textbf{S}^t$ is the value of $\textbf{S}$ in the $t^{\mathrm{th}}$ iteration and 
\begin{align}
\label{eq:lower_bound_Dbi_feb_27_2023}
    \nabla^{\top}_{\textbf{S}}\left(D\left(\textbf{S}^t\right)\right)=-\frac{1}{\mathrm{ln}~2}\sum_{\gidx=1}^K\frac{\sum_{j=1, j\neq \gidx}^K P_j\bar{\textbf{G}}^{\gidx j}}{\sigma^2+\sum_{j=1, j\neq \gidx}^K P_j~\mathrm{tr}\left(\textbf{S}^t\bar{\textbf{G}}^{\gidx j}\right)}.
\end{align} 

By applying \eqref{eq:lower_bound_Dbi_formula_feb_27_2023} to \eqref{eq:finalswitch_SINR_Dbi_Nbi_feb_27_2023}, we obtain a convex SDR problem and solve it using CVX Toolbox in MATLAB~\cite{grant2011matlab}.

\textbf{SFF method:} Heuristic optimization methods (e.g., genetic algorithm and filled function) are popular for optimizing non-convex problems. Unlike the SBO method, a heuristic method directly tackles the non-convex problem in~\eqref{eq:opt_prob_std_RIS_feb_27_2023} without needing approximations for a convex equivalent. This avoids the performance degradation associated with the approximations used in the SBO method. Hence, we utilize the SFF method based on sigmoid filled function~\cite{nassirpour2022power} as our second optimization method in this paper to solve~\eqref{eq:opt_prob_std_RIS_feb_27_2023}. This method includes two steps: local and global search. Prior to explaining the details, we need to define: (i) $\bar{w}(\boldsymbol{\tau}^{\mathrm{s}})\overset{\triangle}=w(\textbf{T}^{\mathrm{s}})$, and (ii) $\left[\boldsymbol{\tau}^{\mathrm{s}}\right]_n\in \{0,1\}^{M \times 1}$ as the $n^{\mathrm{th}}$ neighbor of $\boldsymbol{\tau}^{\mathrm{s}}$, where $||\left[\boldsymbol{\tau}^{\mathrm{s}}\right]_n-\boldsymbol{\tau}^{\mathrm{s}}||^2=1$ and $1\leq n\leq M$. Here, the SFF method first runs the local search with respect to $\bar{w}(\boldsymbol{\tau}^{\mathrm{s}})$ to obtain the local minimum solution for~\eqref{eq:opt_prob_std_RIS_feb_27_2023}. To do so, we start from initial configuration $\boldsymbol{\tau}^{\mathrm{s}}$ and search among its neighbors to find $\boldsymbol{\tau}^{\mathrm{s}*}$ as a new solution that provides $\bar{w}(\boldsymbol{\tau}^{\mathrm{s}*})<\bar{w}(\boldsymbol{\tau}^{\mathrm{s}})$. We note that $\boldsymbol{\tau}^{\mathrm{s}}$ has $M$ neighbors. If $\boldsymbol{\tau}^{\mathrm{s}*}\neq \boldsymbol{\tau}^{\mathrm{s}}$, we set $\boldsymbol{\tau}^{\mathrm{s}}=\boldsymbol{\tau}^{\mathrm{s}*}$, and redo the above process. We define $i^{\text{loc}}$ to indicate the number of iterations that we use the local search process. The search ends if $i^{\text{loc}}$ reaches its maximum value (i.e., $i^{\text{loc}}_{\max}$) or there is no further improvement by running the search (i.e., $\boldsymbol{\tau}^{\mathrm{s}*}=\boldsymbol{\tau}^{\mathrm{s}}$). The details are demonstrated in Algorithm~\ref{algo:Local}. 
\begin{algorithm}
  \caption{Local Search}
  \begin{multicols}{2}
  
  \hspace*{\algorithmicindent} \textbf{Input:} $\boldsymbol{\tau}^{\mathrm{s}}$, $i_{\max}^{\text{loc}}$; \\ 
  \hspace*{\algorithmicindent} \textbf{Output:} $\boldsymbol{\tau}^{\mathrm{s}*}$;
  \begin{algorithmic}[1]
  \label{algo:Local}
  \STATE $i^{\text{loc}}=0$; $\boldsymbol{\tau}^{\mathrm{s}*}=\boldsymbol{\tau}^{\mathrm{s}}$;
  \FOR{ all $ \left[\boldsymbol{\tau}^{\mathrm{s}}\right]_n$ such that $||\left[\boldsymbol{\tau}^{\mathrm{s}}\right]_n-\boldsymbol{\tau}^{\mathrm{s}}||^2=1$}{
  \IF{$\bar{w}(\boldsymbol{\tau}^{\mathrm{s}}_n)< \bar{w}(\boldsymbol{\tau}^{\mathrm{s}*})$}
  \STATE $\boldsymbol{\tau}^{\mathrm{s}*}=\left[\boldsymbol{\tau}^{\mathrm{s}}\right]_n$;
  \ENDIF
  }
  \ENDFOR
  \STATE $i^{\text{loc}} \leftarrow i^{\text{loc}}+1$;
  \IF{$\boldsymbol{\tau}^{\mathrm{s}*}\neq \boldsymbol{\tau}^{\mathrm{s}}$ and $i^{\text{loc}}\leq i_{\max}^{\text{loc}}$}
  \STATE $\boldsymbol{\tau}^{\mathrm{s}}=\boldsymbol{\tau}^{\mathrm{s}*}$;
  \STATE Go to line 2;
  \ELSE
  \STATE $\boldsymbol{\tau}^{\mathrm{s}*}$ is the local minimum solution. 
  \ENDIF
\end{algorithmic}
\end{multicols}
\vspace{-3mm}
\end{algorithm}

The output of local search (i.e., $\boldsymbol{\tau}^{\mathrm{s}*}$) depends on the initial choice of $\boldsymbol{\tau}^{\mathrm{s}}$. Thus, we use global search based on sigmoid filled function to move from $\boldsymbol{\tau}^{\mathrm{s}*}$ to another better solution. Here, we use our filled function, $W_r(\boldsymbol{\tau}^{\mathrm{s}},\boldsymbol{\tau}^{\mathrm{s}*})$, and define an auxiliary optimization problem as below:
\begin{align}
\label{eq:aux_opt}
    &\underset{\boldsymbol{\tau}^{\mathrm{s}}}{\min}~W_r(\boldsymbol{\tau}^{\mathrm{s}},\boldsymbol{\tau}^{\mathrm{s}*}) \nonumber \\
    &\text{s.t.} \quad \boldsymbol{\tau}^{\mathrm{s}},\boldsymbol{\tau}^{\mathrm{s}*}\in \{0,1\}^{M \times 1},
\end{align}
where $r$ denotes the heuristic optimization parameter, $\boldsymbol{\tau}^{\mathrm{s}*}$ is the current local minimum solution,
\begin{align}
\label{eq:filled_func}
    &W_r(\boldsymbol{\tau}^{\mathrm{s}},\boldsymbol{\tau}^{\mathrm{s}*})\nonumber\\
    &~~~~=(1+\frac{1}{1+\eta||\boldsymbol{\tau}^{\mathrm{s}}-\boldsymbol{\tau}^{\mathrm{s}*}||^2})q_r(\bar{w}(\boldsymbol{\tau}^{\mathrm{s}})-\bar{w}(\boldsymbol{\tau}^{\mathrm{s}*})),
\end{align}
\begin{align}
\label{eq:parameters_filled_func}
&q_r(\bar{w}(\boldsymbol{\tau}^{\mathrm{s}})-\bar{w}(\boldsymbol{\tau}^{\mathrm{s}*}))\\ \nonumber
&=
\left\{ \begin{array}{ll}
\vspace{1mm} \bar{w}(\boldsymbol{\tau}^{\mathrm{s}})-\bar{w}(\boldsymbol{\tau}^{\mathrm{s}*})+r , & \bar{w}(\boldsymbol{\tau}^{\mathrm{s}})-\bar{w}(\boldsymbol{\tau}^{\mathrm{s}*})\leq -r, \\
\frac{1}{1+e^{\frac{-6}{r}(\bar{w}(\boldsymbol{\tau}^{\mathrm{s}})-\bar{w}(\boldsymbol{\tau}^{\mathrm{s}*})+r/2)}}, & -r<\bar{w}(\boldsymbol{\tau}^{\mathrm{s}})-\bar{w}(\boldsymbol{\tau}^{\mathrm{s}*})< 0,\\
1, & \bar{w}(\boldsymbol{\tau}^{\mathrm{s}})-\bar{w}(\boldsymbol{\tau}^{\mathrm{s}*})\geq 0,
\end{array} \right.
\end{align}
\begin{align}
    & \eta=
\left\{ \begin{array}{ll}
\vspace{1mm} 0, & \bar{w}(\boldsymbol{\tau}^{\mathrm{s}})-\bar{w}(\boldsymbol{\tau}^{\mathrm{s}*})\leq -r, \\
1, & \text{otherwise}.
\end{array} \right.
\end{align}

In this part, we use $i$ and $\boldsymbol{\tau}^{\mathrm{s}}_i$ to denote the $i^{\mathrm{th}}$ iteration of global search and the solution in the $i^{\mathrm{th}}$ iteration, respectively. Here, we start from an initial random solution,  $\boldsymbol{\tau}^{\mathrm{s}}_0$, and then apply Algorithm~\ref{algo:Local} to~\eqref{eq:opt_prob_std_RIS_feb_27_2023} to get $\boldsymbol{\tau}^{\mathrm{s}*}_0$. After finding $\boldsymbol{\tau}^{\mathrm{s}*}_0$, we run local search starting from $\boldsymbol{\tau}^{\mathrm{s}*}_0$ with respect to~\eqref{eq:aux_opt} to attain $\bar{\boldsymbol{\tau}}^{\mathrm{s}*}_0$. Next, we assume $\boldsymbol{\tau}^{\mathrm{s}}_1=\bar{\boldsymbol{\tau}}^{\mathrm{s}*}_0$. If $\bar{w}(\boldsymbol{\tau}^{\mathrm{s}}_1)<\bar{w}(\boldsymbol{\tau}^{\mathrm{s}*}_0)$, we consider $\boldsymbol{\tau}^{\mathrm{s}}_1$ as a new solution and redo the above procedure. Else, we apply other neighbors of $\boldsymbol{\tau}^{\mathrm{s}*}_0$ to~\eqref{eq:aux_opt} to obtain a new $\bar{\boldsymbol{\tau}}^{\mathrm{s}*}_0$ and repeat the above procedure. If neither $\boldsymbol{\tau}^{\mathrm{s}*}_0$ nor its neighbors can lead to a new solution, we reduce $r$ (e.g., $r=r/10$) to find a new filled function, and thus new $\bar{\boldsymbol{\tau}}^{\mathrm{s}*}_0$ and then repeat the above search. Moreover, we use $\gamma \in \mathbb{N}$ to indicate that after $\gamma$ times of finding
the local minimizer of~\eqref{eq:aux_opt}, we
run local search for~\eqref{eq:opt_prob_std_RIS_feb_27_2023}. The search stops if $r< \epsilon, \epsilon\ll 1$ or $i^{\text{filled}}$, the number of times that we use the filled function, reaches its maximum value, $i^{\text{filled}}_{\max}$. In this case, we say that the current solution is an approximation of the global minimum solution.


\begin{algorithm}
  \caption{Global Search}
  \begin{multicols}{2}
  \hspace*{\algorithmicindent} \textbf{Input:} $\boldsymbol{\tau}^{\mathrm{s}}_0, r, \gamma, \epsilon, i_{\max}^{\text{loc}}, i_{\max}^{\text{filled}}$; 
  \hspace*{\algorithmicindent}\hspace{-0.4mm} \textbf{Output:} $\boldsymbol{\tau}^{\mathrm{s}**}$;
  \begin{algorithmic}[1]
  \label{algo:global}
  \STATE $i=0$; $r_0=r$; $\bar{\gamma}=\gamma$;
  \STATE $i^{\text{filled}}=0$;
  \STATE $\boldsymbol{\tau}^{\mathrm{s}**}=\boldsymbol{\tau}^{\mathrm{s}}_0$;
  \IF{$i+1\geq \bar{\gamma}$}
  \STATE $\bar{\gamma}\leftarrow\bar{\gamma}+\gamma$;
  \STATE Apply Algorithm~\ref{algo:Local} to $\boldsymbol{\tau}^{\mathrm{s}}_{i}$ and \eqref{eq:opt_prob_std_RIS_feb_27_2023} to get $\boldsymbol{\tau}^{\mathrm{s}*}_{i}$;
  \ELSE
  \STATE  $\boldsymbol{\tau}^{\mathrm{s}*}_{i}=\boldsymbol{\tau}^{\mathrm{s}}_{i}$;
  \ENDIF
  
  \IF{$i=0$ or $\bar{w}(\boldsymbol{\tau}^{\mathrm{s}*}_{i})<\bar{w}(\boldsymbol{\tau}^{\mathrm{s}**})$ }
      \STATE $\boldsymbol{\tau}^{\mathrm{s}**}=\boldsymbol{\tau}^{\mathrm{s}*}_{i}$; $r=r_0$; 
      \STATE $m=1$;
      \STATE Apply Algorithm~\ref{algo:Local} to $\boldsymbol{\tau}^{\mathrm{s}*}_{i}$ and \eqref{eq:aux_opt} to obtain $\bar{\boldsymbol{\tau}}^{\mathrm{s}*}_{i}$;
      \STATE $i^{\text{filled}} \leftarrow i^{\text{filled}}+1$
      \IF{$m=1$}
      \STATE $i\leftarrow i+1$;
      \STATE $\boldsymbol{\tau}^{\mathrm{s}}_{i}=\bar{\boldsymbol{\tau}}^{\mathrm{s}*}_{i-1}$;
      \ELSE
      \STATE $\boldsymbol{\tau}^{\mathrm{s}}_{i}=\bar{\boldsymbol{\tau}}^{\mathrm{s}*}_{i}$;
      \ENDIF
      \STATE Go to line 4;
  \ENDIF
  \IF{$m \leq M$}
      \STATE $\boldsymbol{\tau}^{\mathrm{s}*}_{i}=$ the $m^{\mathrm{th}}$ neighbor of $\boldsymbol{\tau}^{\mathrm{s}*}_{i-1}$;
      \STATE $m\leftarrow m+1$;
      \STATE Go to line 12;
  \ELSE
      \IF{$r<\epsilon$ or $i^{\text{filled}}>i_{\max}^{\text{filled}}$ }
           \STATE $\boldsymbol{\tau}^{\mathrm{s}**}$ is the global minimum solution.
          \ELSE
          \STATE $r \leftarrow \frac{r}{10}$;
          \STATE $i \leftarrow i-1$;
          \STATE Go to line 11;   
      \ENDIF
  \ENDIF
    \end{algorithmic}
    \end{multicols}
    \vspace{-3mm}
\end{algorithm}

In Section~\ref{Section:Num_results}, we will show that the SFF method outperforms the SBO method using S-RIS. 

In the second scenario based on \newwRIS, we need to update~\eqref{eq:opt_prob_std_RIS_feb_27_2023} using $\textbf{T}=\textbf{T}^{\mathrm{i}}$. Here, we use the SFF method by following Algorithms~\ref{algo:Local} and \ref{algo:global} since the filled function-based optimization methods solve a general non-convex integer programming problem as long as selecting a possible configuration from a discrete set is the only constraint~\cite{ng2005discrete}. On the contrary, the SBO method can not be directly applied to \newRIS due to the additional non-convexity that arises from dependence of $\bar{\mathcal{M}}^{[m]}$ on the current solution. This necessitates an additional approximation, which further degrades its performance. As a result, in the next section, we only utilize the SFF method for optimizing the I-RIS elements.

\section{Numerical Analysis}
\label{Section:Num_results}
In this section, we evaluate the effectiveness of our proposed RIS structures. We assume a multi-user RIS-assisted network with one RIS and $K$ transmitter-receiver pairs, where transmitters and RIS are fixed-location nodes and receivers are located randomly across an area with $100 \mathrm{m}\times 100\mathrm{m}$ dimensions. The simulation parameters are shown in Table~\ref{table:sim_param}.
\vspace{-3mm}
\begin{table}[h]
\caption{Simulation parameters}
\fontsize{9pt}{12pt}\selectfont
\centering
\begin{tabular}{|c|c|}
\hline
\textbf{Parameters} & \textbf{Value} \\ 
\hline 
$\#$ of transmitter-receiver pairs & $K=4$\\ Location of transmitters $(\mathrm{10m})$ & $(5,0),(0,5),(10,5),(5,10)$\\
Location of RIS $(\mathrm{m})$& $(50,50)$\\
Transmit power $(\mathrm{dBm})$ & $P_j=40, 1 \leq j \leq K$\\
Noise power $(\mathrm{dBm})$  & $\sigma^2=-80$\\
Path loss exponent & $\alpha_1=2,\alpha_2=2.1,$\\
Signal loss at $1\mathrm{m}$ $(\mathrm{dB})$ & $C_0=-30$\\
Wavelength $(\mathrm{m})$& $\lambda=0.125$\\
Rician factor & $\kappa=2$\\
SFF parameters & $r=10, \gamma=10, \epsilon=0.01$\\
 & $i_{\max}^{\text{loc}}=M, i_{\max}^{\text{filled}}=8(M+1)$\\
SBO parameter & $I_{\mathrm{tr}}=5$\\
\hline
\end{tabular}
\label{table:sim_param}
\vspace{-6mm}
\end{table}

\begin{figure}[ht]
  \centering
  \includegraphics[trim = 0mm 0mm 0mm 0mm, clip, scale=5, width=0.62\linewidth, draft=false]{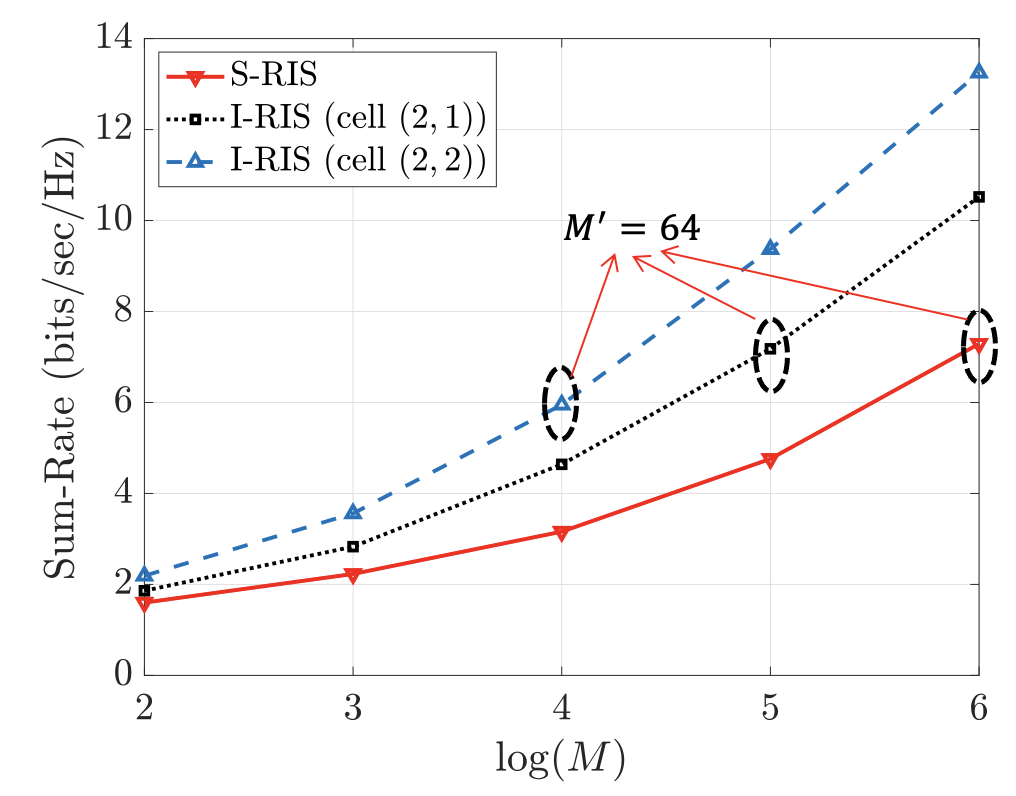}
  \vspace{-4mm}
  \caption{Sum-rate against $\log(M)$ using S-RIS, I-RIS $(2,1)$, and I-RIS $(2,2)$.}\label{Fig:S_RIS_vs_I_RIS}
  \vspace{-2mm}
\end{figure}

\textbf{I-RIS vs. S-RIS:} In this part, we present a sum-rate comparison between S-RIS and I-RIS using the SFF method. Fig.~\ref{Fig:S_RIS_vs_I_RIS} shows sum-rate versus $\log(M)$ using S-RIS and two I-RIS cases with cell size $(2,1)$ and $(2,2)$ where $M \in \{4,8,16,32,64\}$. Based on Fig.~\ref{Fig:S_RIS_vs_I_RIS}, I-RIS enables multi-user scaling by offering a higher gain when $M$ is fixed. For instance, with $M=64$, I-RIS $(2,1)$ and  I-RIS $(2,2)$ offer $44\%$ and $81\%$ improvement compared to S-RIS, respectively. This benefit comes from using more electronic circuits (power splitters and additional RF switches) to share the incident signal powers in I-RIS. We note that it is possible to create different versions of I-RIS by changing the cell size, and as the cell size increases, we get a higher gain. 

Moreover, Fig.~\ref{Fig:S_RIS_vs_I_RIS} depicts that I-RIS requires fewer elements than S-RIS to provide the same sum-rate. For example, to achieve a sum-rate of $6$ bits/sec/Hz, we need $64$ elements with S-RIS, whereas we require $32$ and $16$ elements to get the same performance with I-RIS $(2,1)$ and  I-RIS $(2,2)$, respectively. This occurs since I-RIS needs fewer elements than S-RIS to obtain the same DoC. We use $M^{\prime}$ to denote the DoC, where $M^{\prime}=M\times c\times d$ with I-RIS, whereas in an S-RIS, we have $M^{\prime}=M$. Fig.~\ref{Fig:S_RIS_vs_I_RIS} shows that S-RIS and I-RIS provide almost identical sum-rates if they have the same value of $M^{\prime}$.

\begin{figure}[ht]
\centering
  \includegraphics[trim = 0mm 0mm 0mm 0mm, clip, scale=5, width=0.64\linewidth, draft=false]{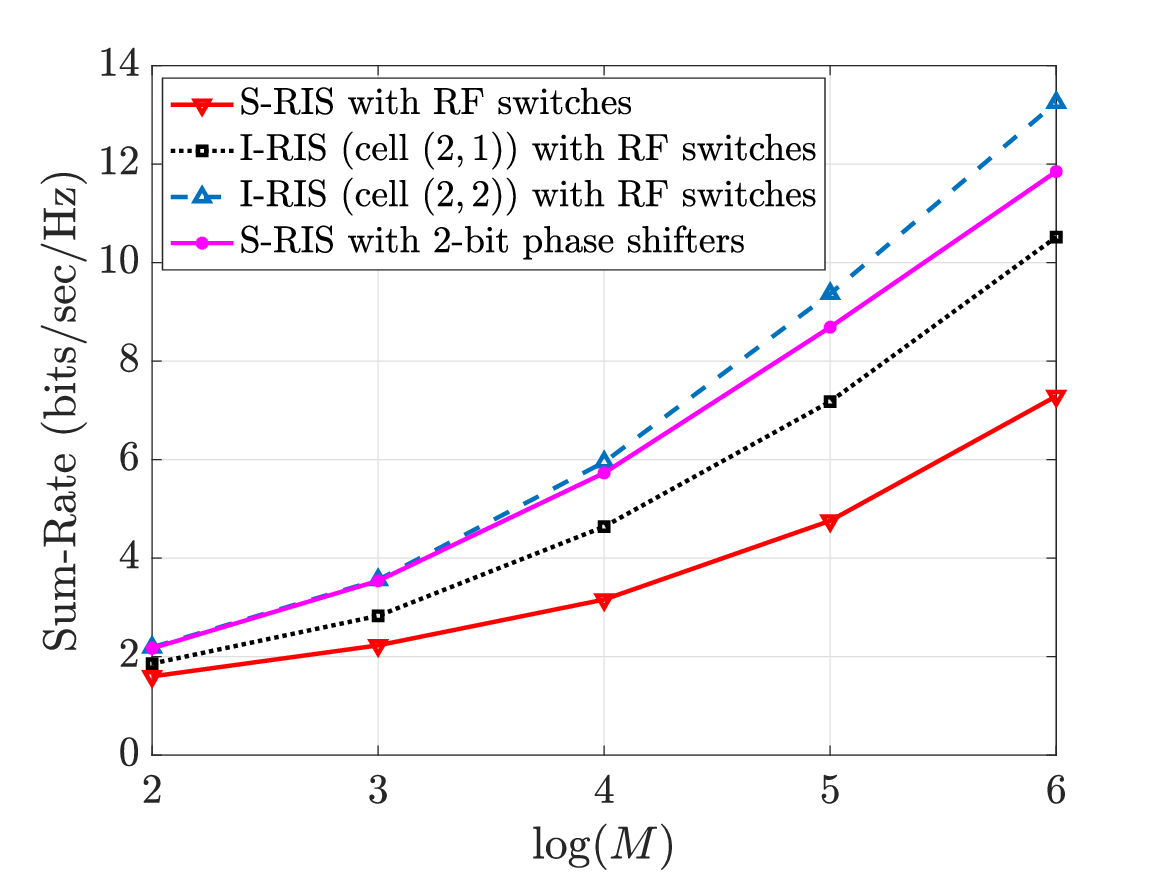}
  \caption{A comparison between a switch-based S-RIS, two cases of switch-based I-RIS, and a PS-based S-RIS using the SFF method. 
  }\label{Fig:sw_vs_PS_I_RIS}
  \vspace{-6mm}
\end{figure}
\textbf{Switch against phase shifter:} In Fig.~\ref{Fig:sw_vs_PS_I_RIS}, we compare the performance of our proposed switch-based RIS structures with an S-RIS using $2$-bit PSs. According to Fig.~\ref{Fig:sw_vs_PS_I_RIS}, our switch-based I-RIS $(2,2)$ has a higher gain due to its higher DoC ($M^{\prime}=M\times b$ with an S-RIS using $b$-bit PSs). Additionally, Fig.~\ref{Fig:sw_vs_PS_I_RIS} shows that if our switch-based I-RIS and the S-RIS with PSs have the same DoC (i.e., I-RIS $(2,1)$ and S-RIS with $2$-bit PSs), the latter provides a higher gain since it changes the phase of the incident signals, while the switch-based RIS can either block or reflect the incident signals.

\begin{figure}[ht]
\vspace{-4mm}
\centering
  \includegraphics[trim = 0mm 0mm 0mm 0mm, clip, scale=5, width=0.64\linewidth, draft=false]{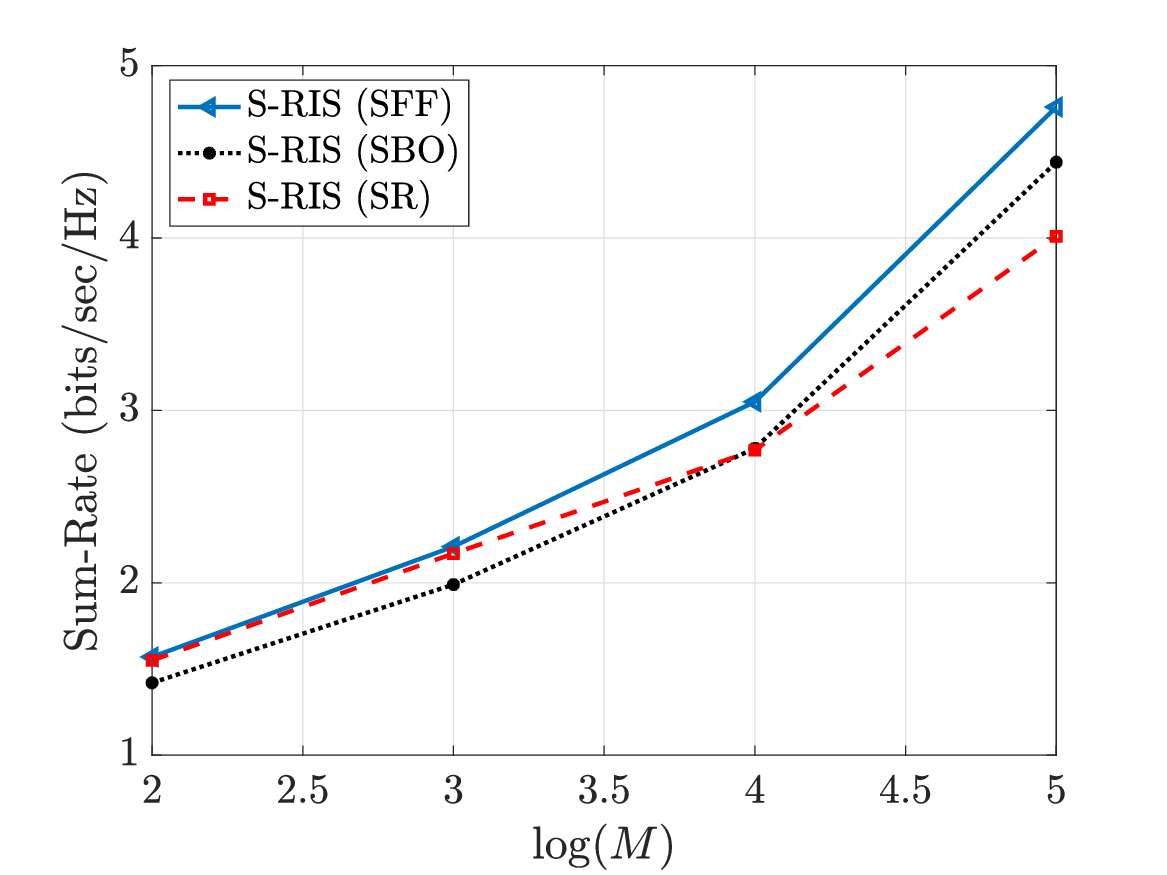}
  \vspace{-4mm}
  \caption{A Sum-rate comparison between the SFF, SBO, and SR methods. 
  }\label{Fig:optimization_comparison}
  \vspace{-4mm}
\end{figure}

\textbf{Performance of the optimization methods:} In this part, we evaluate the performance of our proposed optimization methods (the SFF and SBO methods) and the SR method in terms of sum-rate and complexity. Fig.~\ref{Fig:optimization_comparison} depicts sum-rate versus $\log(M)$ using S-RIS. It shows that the SFF method offers a higher sum-rate than the others since the SFF method directly targets the non-convex problem to
optimize the RIS elements, while the SBO method incorporates some approximations in the optimization process and solves an equivalent convex problem. Also, the SR method finds a local optimum solution, whereas the SFF method uses a filled function to move from one local optimum solution to another better one. Furthermore, the complexity of the SFF method is equal to $\mathcal{O}(2\log_{10}(r/\epsilon)M^5)$~\cite{nassirpour2022control}, which is lower than the complexity of  
the SBO method,
$\mathcal{O}(I_{\mathrm{tr}}(M+1)^3M^3)$~\cite{bomze2010interior}, and higher than the complexity of the SR method, which is $\mathcal{O}(M^4)$~\cite{wu2019beamforming}.

\section{Conclusion}
\label{Section:conclusion}
In this paper, we introduced a new compact RIS structure with interconnected elements, I-RIS, suited for space-limited nodes to assist multi-user networks. We used I-RIS to enable multi-user scaling and do so with fewer elements compared with proper designs. We considered binary RF switches as RIS elements to mitigate the practical issues related to phase shifters and showed that I-RIS outperforms an S-RIS using phase shifters. We proposed the SFF and SBO methods to obtain the optimal RIS configuration. Our findings showed that the SFF method provides a higher sum-rate than the SBO and SR methods, and its complexity is higher than the SR method and lower than the SBO method. In future works, we would like to use multiple I-RISs to improve the performance of multi-user networks under realistic channel models.

\bibliographystyle{IEEEtran}
\bibliography{refs}

\end{document}